\documentclass[11pt,letterpaper]{llncs}
\usepackage{amsmath}
\usepackage{amssymb}
\usepackage{graphicx}
\usepackage{marvosym}
\usepackage{url}
\usepackage{fancyhdr}
\usepackage{verbatim}

\usepackage{booktabs}
\usepackage{afterpage}
\usepackage{microtype}
\usepackage{subfigure}

\usepackage{geometry}
\newcommand{\keywords}[1]{\par\addvspace\baselineskip
\noindent\keywordname\enspace\ignorespaces#1}

\fancypagestyle{firstpage}{\fancyhf{}
\setcounter{page}{1}
\fancyhead[C]{\small{International Journal of Software Engineering $ \& $ Applications (IJSEA), Vol.11,No.4,July 2020}}
\fancyfoot[L]{DOI: 10.5121/ijsea.2020.11402}
\rfoot{\thepage}
}

\AtBeginDocument{%
\raggedbottom
\hbadness=99999
\hfuzz=9999pt
  \providecommand\BibTeX{{%
    \normalfont B\kern-0.5em{\scshape i\kern-0.25em b}\kern-0.8em\TeX}}}
\begin{document}

%\mainmatter  % start of an individual contribution

\title{\LARGE{How do Agile Software Startups deal with uncertainties by Covid-19 pandemic?}}

%\titlerunning{Lecture Notes in Computer Science: Authors' Instructions}


\author{\large{Rafael da Camara \inst{1,2} \and Marcelo Marinho \inst{1} \and Suzana Sampaio \inst{1} \and Saulo Cadete \inst{2} }}
\institute{\large{Federal Rural University of Pernambuco, Recife, Pernambuco \\
              Department of Computer Science} \\
              %\email{(rafael.camara;marcelo.marinho;suzana.sampaio)@ufrpe.br}\\ 
              \and
              \large{Di2win, Recife, PE, Brazil} \\
              %\email{saulo@di2win.com}
              }

%\author{Alfred Hofmann%
%\thanks{Please note that the LNCS Editorial assumes that all authors have used
%the western naming convention, with given names preceding surnames. This determines
%the structure of the names in the running heads and the author index.}%
%\and Ursula Barth\and Ingrid Haas\and Frank Holzwarth\and\\
%Anna Kramer\and Leonie Kunz\and Christine Rei\ss\and\\
%Nicole Sator\and Erika Siebert-Cole\and Peter Stra\ss er}

%\institute{Springer-Verlag, Computer Science Editorial,\\
%Tiergartenstr. 17, 69121 Heidelberg, Germany\\
%\url{http://www.springer.com/lncs}}


%\toctitle{Lecture Notes in Computer Science}
%\tocauthor{Authors' Instructions}

\maketitle

\thispagestyle{firstpage}

\begin{abstract}
The dissipation of severe acute respiratory syndrome coronavirus 2 (SARS-CoV-2) has already taken on pandemic proportions, affecting over 100 countries in a couple of weeks. The evolution of the disease and its economic impact is highly uncertain, which brings challenges for newly created software companies. Software startups are companies that create innovative software products and services in a dynamic and fast-growing market. Agile Software Methods aims to enable startups in responding to uncertainty caused by Covid-19. This paper investigates the impact of Covid-19 in a real software startup context to understand how they have reacted against uncertainties caused by Covid-19. As a research methodology, action research within Di2Win, a Brazilian software startup, has been applied. The study was carried out throughout six sprints, during the quarantine. Practices employed to mitigate threats while simultaneously allowing teams to remain open to opportunities and challenges are detailed. This paper shares lessons learned that could help agile software startups improve their way of work in an uncertain environment caused by the Covid-19 pandemic.
\end{abstract}

\section{Introduction} \label{intro}

The pandemic is already reshaping the way companies work, also in software engineering, and generates changes in processes, methods, use of collaboration tools, etc. Other effects are likely to be seen in the coming years.

One of the immediate ways through which COVID-19 has impacted most companies is the social distancing. For companies that had already adopted virtual teams, this is just another typical day. Although for others, it is likely to have a temporary impact on projects, team productivity and collaboration.

Since agile methods started to be applied to software development, software teams and managers have demanded additional capabilities to achieve better results at the business level \cite{dingsoyr2012decade}. For example, agile teams look for the ability to sense and respond to changes in a coordinated way across the company. %Fitzgerald and Stol \cite{fitzgerald2017continuous} note that business and development disconnect in their study on continuous software engineering. Hence, it is reasonable to consider that challenges also emerge as organizations scale, which in turn requires new knowledge and approaches to manage them.
%The Agile philosophy impacts upon management, collaboration, control, new skills, training or recruitment styles \cite{hohl2018back}. Agile methods are built around trained and self-organized teams, with a strong focus on collaboration and communication supported by various agile practices. 
Agility co-exists with uncertainty which is unavoidable in the teams using agile methods. Uncertainty needs to be embraced. Embraced uncertainty is manifested firstly as the probability to change directions. All the teams believe that the iterative nature of their process gives them more possibilities to change directions when needed \cite{conboy2009understanding}.

Based on the presented context, an empirical study of a startup which allowed us to illustrate our research objectives in a very tangible way was conducted. In this study, we aim to present the actions taken in the startup to manage uncertainties that emerged from the Covid-19 pandemic. The main research question underlying our study is: How do software development startups utilizing agile approach uncertainties caused by Covid-19s?

Rather than working in a remote office or well-appointed home office, some people are working in extempore in bedrooms, at kitchen tables, and on sofas while partners, children, siblings, parents, roommates, and pets distract them \cite{donnelly2015disrupted}. This paper does not focus on Distributed Software Development (DSD) \cite{Uncertainty&GTM2018}. But ``working from home'', unexpectedly, during an unprecedented crisis (remote way of working). So, DSD benefits of working from home do not apply . The aim of this paper is to present a transformation of an agile co-located startup into a working from home approach. Furthermore, this study also summarizes a set of lessons learned that can be used by other startups. 

This paper is organized as follows: in Section \ref{sec:background} we introduce the background of the problem, and define our research questions. Section \ref{sec:method} describes the method used. In Section \ref{sec:results} we present the results, and discuss their implications and limitations; Section \ref{sec:conclusions} presents conclusions and future directions.

\section{Background} \label{sec:background}

\subsection{Agile}  \label{sec:agile}

The majority of companies, over 90\% according to a recent estimate \cite{versionOne} are adopting an agile approach for software development. Becoming agile often goes along with fundamental changes that are facing a lot of uncertainties \cite{marinho2018uncertainty}. Regardless the industry or the company’s motivation – the dilemma is always the same: while the decision of doing agile is made easily, actually becoming agile is not \cite{hohl2018back}. Many companies claim to be “agile”, although they often neglect that agile has two components to consider: technical and cultural agility \cite{hohl2018back}. The neglect of cultural agility may lead to the dissatisfaction of the people involved due to uncertainties in software development activities.

% Sugestão de corte:
%Kruchten \cite{kruchten2013contextualizing} defines agility as ``the ability of an organization to react to changes in its environment faster than the rate of those changes''. This definition uses the ultimate goal of being agile for business, rather than defining agility by a set of labelled practices (for example, you are agile when running XP \cite {beck2000extreme}, Scrum \cite {ScrumBook} or Lean \cite{poppendieck2007implementing}) or by a set of properties defined as opposed to another set - the agile manifest \cite{fowler2001agile}. This definition is not far from Conboy's, which is addressed in his research on the literature on agile process development \cite {conboy2011people}.

\subsection{Uncertainty} \label{sec:uncertainty}

An \emph{uncertainty} is a risk with unknown probability. A focus on uncertainty rather than risk has been suggested to improve project management by ``providing an important different perspective, including, but not limited to, an enhanced focus on opportunity management'' \cite{ward2003transforming}.

Following product development researchers \cite{miller2001understanding,ward2003transforming,marinho2018uncertainty}, we rely on the uncertainty conceptualization that encompasses the impact of ``anything that matters'', defining it as the lack of certainty that leads to a situation in which ``potential outcomes and causal forces are not fully understood'' \cite{miller2001understanding}. Explicitly, in line with Marinho \cite{marinho2018uncertainty} we define uncertainty as incomplete information that bears the potential for positive or negative consequences of high impact on project objectives. In other words, uncertainty is the discrepancy between the information that is available and the information that is required (but not yet available) toward reaching a goal \cite{marinho2018uncertainty}.

To Marinho et. al \cite{marinho2014systematic,marinho2018uncertainty}, the key elements of any software development project are the identification of potential uncertainty sources and the ability to respond to changes during the software lifecycle. In this perspective, \cite{marinho2018uncertainty} classifies the uncertainty sources in a project as: 
%adição
technological, market, environmental and socio-human.

% Sugestão de corte:
\begin{comment} 

\begin{itemize}
    \item Technological uncertainty: this dimension is associated with uncertainty about the knowledge and application that the project makes use of technology; 
    \item Market uncertainty: this dimension refers to the degree of novelty of the product, service or outcome of the project; 
    \item Environmental uncertainty: this dimension is directly determined by the complexity and dynamic environment in which organizations are inserted; 
    \item Socio-human uncertainty: this dimension is determined by the relationships among people of an organization. These relationships have cognitive factors and are intrinsically related to how people perceive, learn, think, and assimilate the information within the organization. 
\end{itemize}

\end{comment}

Different approaches are important because uncertainty is not inherently good or bad. Whereas uncertainty affects decisions that can cause projects and even entire companies to fail, uncertainty is also a vital predecessor of innovation \cite{marinho2018uncertainty}. Uncertainty can be seen in any part of the startup process. Uncertainty is not just an uncomfortable match that the founders cannot avoid; it is an integral part of what allows startups to be successful.

\subsection{Startups} \label{sec:startups}

Startups are organizations devised to create new products or services under conditions of uncertainty, which seek rapid growth and repeatable, profitable and scalable business models \cite{ries2011lean}. Software startups have as main focus the development of innovative products or services, using software, from which the commercial value is created. Although software startups share common characteristics with other types of startups, such as scarcity of resources and lack of operational history \cite{sutton2000role}, they are often accompanied by the wave of technological changes that often occur in the software industry, such as new computing technologies and network \cite{pantiuchina2017software}.

% Sugestão de corte:
%Software startups are essential for the economy due to their potential to create jobs and add value to domestic products on a global scale \cite{tripathi2018anatomy}. New software ventures such as Facebook, Linkedin, Spotify, Pinterest, Instagram, and Dropbox, to name a few, are examples of startups that evolved into successful businesses \cite{berg2018software}. Inspired by success stories, a large number of software businesses are created every day \cite{pantiuchina2017software,tripathi2018anatomy}. However, the great majority of these companies fail within two years from their creation \cite{crowne2002software}.

As the ability to accommodate frequent changes is essential in the startup context, agile methods have been considered the most suitable process model for software startups, as they allow them to adopt changes and enable development to adapt to business strategies \cite{pantiuchina2017software}. Frequent delivery with an iterative and incremental approach, used in agile philosophy, reduces waiting time, from idea conception to production and the market.

However, software startups are always under enormous pressure from time to market and currently need to address uncertainties quickly due to the new coronavirus.

\subsection{Novel Covid-19} \label{sec:covid-19}

On January 30, 2020, the World Health Organization (WHO) declared the spread of a new type of coronavirus, SARS-CoV-2 \cite{gorbalenya2020severe}, as a public health emergency of international interest \cite{world2008international}. This virus was first identified in patients with pneumonia in the city of Wuhan, in the province of Hubei, China, in December 2019 \cite{chen2020rna} and, due to its rapid transmissibility, it gained prominence in the world scientific community \cite{chen2020rna}.

%Sugestão de corte:
%After the emergence of the virus in China, the first cases started to show up in Europe between January to February \cite{coronaFirstCasesEU}, 2020. By the beginning of March, European governments started more drastic policies to prevent the spread of the virus. Italy started the lockdown at the Lombardy region, on March 7Th to contain the spread \cite{guardianItaly}, and thirteen days later the government established nationwide lockdown, closing parks, non-essential factories and banning outdoor activities \cite{nyTimesItaly}. Spain ordered the nationwide lockdown on March 14Th, 2020 letting only essential services be open such as supermarkets, pharmacies, and gas stations \cite{guardianSpain}. France established its nationwide lockdown on March 17Th, 2020 for 30 days, closing bars, restaurants, schools, and universities \cite{guardianFrance}. Soon, Germany imposed its lockdown in the middle of March for fifteen days but extended it for more fifteen days due to the growing spread of the virus \cite{bloombergGermany}. By the time the number of COVID-19 was growing in Europe, the first case was confirmed in Brazil on February 26Th, 2020 \cite{aljazeeraBrazil}.

The software development community is facing many uncertainties due to the work environment changes caused by the pandemic. According to Paul Ralph \textit{et al.,} \cite{Ralph2020Pandemic}, the home office ergonomics, the distractions caused by people with whom the team members live, the absence of fitness facilities and the fear of pandemic can truly affect well being and productivity of the team members.

As a worldwide virus, Covid-19 is affecting everyone, although some companies are particularly affected by the virus. Small software development companies and Startups that do not have the founds or the maturity to deal with the uncertainties related to such change are the most impacted. Thus, given the uncertainties arising by Covid-19 in agile startups, this study has the following research question:

\emph{How do agile software development Startups approach uncertainties caused by Covid-19s?}

\section{Method}  \label{sec:method}

Several empirical evaluation approaches can be identified, including case studies, ethnographies, experiments, surveys, and action research \cite{baskerville1999investigating}. Among these, action research appears to be an essential and valid instrument for evaluating the impact of uncertainties by Covid-19 within an Agile Software Startup context. Action research is an empirical research methodology through which researchers are trying to figure a real-world problem meanwhile studying the experience of solving the problem. 

%Sugetão de corte:
%Easterbrook et al. \cite{easterbrook2008selecting} argue that a lot of software engineering research is disguised action research. Indeed, many key ideas in software engineering have initially been developed by experiencing them in real development projects and reports on experiences. In this sense, Dos Santos and Travassos \cite{dos2011action} describes the cooperative systems development as an ideal action research for empirical software engineering. By adopting the action research structure more explicitly, the design and evaluation of such research may become stricter.

The action research process can be defined as several learning cycles consisting of predefined stages. This research undertakes three cycles of action research. We conduct action research following the protocol proposed by Baskerville et al. \cite{baskerville1999investigating}.

%Sugestão de corte:

\begin{comment}

In general, action research usually includes the following activities:

\begin{enumerate}
    \item \textbf{Diagnosis:} this phase corresponds to the identification of primary problems triggering the desire for a change in an organization;
    \item  \textbf{Action planning:} this is the phase where you plan the actions to address the issues that are identified in the diagnosis phase; 
    \item \textbf{Action taking:} this is the phase in which the actions that are scheduled in the action planning phase are executed;
    \item  \textbf{Evaluation:} this is the phase in which the evaluation of the action taking is conducted.
    \item \textbf{Reflection and Learning:} in this phase, the assessment of the action taking is undertaken. Here the researchers evaluate whether the theoretical effects of the actions are fullfilled or not.
\end{enumerate}

\end{comment}

The action research's main characteristic is the involvement of the practitioners as both subjects and co-researchers. Similarly, our action research team consisted of two internal researchers and two external researchers (hereafter referred to as ``research team''). Workshops and meetings were held at which academic theory and professional practice were discussed to iteratively propose actions to address the many uncertainties and challenges that come with the pandemic. To quickly support the company, the research team focused on weekly sprints with an application of an action research cycle every two sprints. We performed an action research protocol (\url{https://bit.ly/2W13WVe}) and the dataset is available at \cite{dataset}.

%\url{https://bit.ly/2O7M5rj}

\subsection{Company Context} \label{sec:company}
Our study was conducted inside the operation of Di2Win \footnote{www.di2win.com}. Di2win is a startup based in Brazil, but with partners around the world. The company headquarters is located in Recife, a city in the Northeast of Brazil. The startup was built in 2018 and it emerged from the experience of a group of entrepreneurs, researchers, and tech experts with more than 25 years of research and development of solutions based on artificial intelligence. Following the maturity of artificial intelligence technologies, and the market capacity to demand and absorb them, the startup saw the need to provide a digital transformation to all economic sectors. The first step was to connect concepts from process automation, and robotic to build a platform capable of adapting to business process needs of any company, improving the process with artificial intelligence, and digitalizing all its steps.

% Sugestão de corte:
%Di2win develops products and solutions with state of the art technology combined with artificial intelligence, process automation and apply them to their agile process transformation framework. Today, the startup has more than 20 employees in Recife and more than 10 employees around the globe conducting projects from all types, such as banking, insurance, education, telecommunications, transport, social purpose, and others. Most of the projects are developed using agile practices from Scrum \cite{ScrumBook} and Kanban \cite{brechner2015agile}. Due to the pandemic, the startup aims to adapt these practices to remote work.

\subsection{Project Context} \label{sec:project}

% Sugestão de corte:
%The studied project has been developed since September 2019 and it was first presented as proof of concept for the customer in the same month. After the negotiations and the agreement with the customer, the first version of the project started to be developed in November 2019. 
The project is an insurance onboarding platform for automobiles that communicates with a legacy ERP (Enterprise Resource Planning) through two Web-applications that works as an interface. The project covers all the processes of automobile insurance, since the quotation, the proposal creation and effectuation, and policy emission. All the process phases are made over the web-applications by both the end-users and the operation users, and all the communication between the web-applications and the ERP is made over robotic process automation.

% Sugestão de corte:
%The first version of the project was conducted with a process owner of the customer that is a specialist in other insurance products of the company, but that does not have an overall domain of the car insurance market. Due to this fact, some of the requirements were misunderstood by the team and were consequently implemented differently. However, by March 2020, the customer and the team decided to start the development of the second version of the project, now with a process owner specialized in the car insurance market. The beginning of this phase was marked by the survey of requirements with the specialist, and when a minimum backlog was ready, the team started the development on March 19th, 2020.

The studied team is formed by one scrum master, one product owner, one technical leader, and six developers. We gave each one an ID to preserve their identity as shown in Table \ref{r:tab1}. The study presents actions that impact all team members and the other employees of the company, but the focus of the study was given to the software development members.

All Di2win project teams use a combination of agile principles, practices, and events from frameworks and models such as Scrum \cite{ScrumBook} and Kanban \cite{brechner2015agile} in their development process. 
%Sugestão de corte:
%The most common agile events and activities held by the teams are sprints, daily stand-up meetings, sprint planning, sprint reviews, retrospectives meetings, product and sprint backlog management, kanban board, continuous delivery, and others. 

\begin{table}
\footnotesize
\centering
\caption{Developer ids and roles description}
\label{r:tab1}
\begin{tabular}{@{}cll@{}} 
\toprule
\textbf{Developer ID} & \multicolumn{1}{c}{\textbf{Role}} & Target Application Domain     \\ \midrule
P1                    & Technical leader                  & \multicolumn{1}{c}{}          \\
P2                    & Frontend developer                &                               \\
P3                    & Frontend Developer                &                               \\
P4                    & Automation Developer              &                               \\
P5                    & Backend Developer                 & Insurance onboarding platform \\
P6                    & Scrum Master                      &                               \\
P7                    & Product Owner                     &                               \\
P8                    & Requirement Developer             &                               \\
P9                    & Requirement Developer             &                               \\ \bottomrule
\end{tabular}
\end{table}

\section{Results}  \label{sec:results}

\subsection{Diagnosis} \label{sec:diagnosis}

\subsubsection{Problem Description} \label{sec:problem}

Before the Covid-19 pandemic, Di2win used to have most of its operation in a co-located office situated in Recife. However, with the Covid-19 propagation, the possibility of the disease high spread in Recife became evident. Therefore, an immediate change of mindset and the necessity to adapt processes, guidelines and even the workplace became urgent.

Suddenly, the situation changed overnight, the cases of Covid-19 started to appear in Recife on March 12th, 2020.
%\cite{g1Recife}.
On March 16th, 2020, the state government decreed the suspension of classes from private and public schools and Universities %\cite{decree16/03}. 
Further ahead, on March 20th, 2020, the state government decreed the suspension of the activities of factories, commercial, and service provision establishments considered non-essentials.
%\cite{decree20/03}. 
However, before all these government  decrees, the research team started to discuss the situation and possible impact on the company's projects. In the meantime, during the discussions, the pandemic raised many uncertainties regarding project contracts, customers' response to the pandemic, team's capacity to work remotely (from home), infrastructure that needed to be provided, and suppliers' work. 

The possibility of a lockdown demanded the team to adapt the agile practices co-located to a work from home, and consequently manage the uncertainties that could arise from it. The biggest challenges were (i) to maintain the productivity of the teams, (ii) to define the tools needed to manage the remote work, (iii) to align expectations with the clients', (iv) to continue delivering value through cycles, (v) to maintain the employee welfare, (vi) to provide the necessary infrastructure for all employees, and (vii) to coordinate the development process.

Due to the chaotic situation generated by the pandemic and by the will of Di2win to keep the quality of its software development process, it was proposed by the first author (the technical leader from the studied project) to conduct an action research to present, investigate, and evaluate practices and tools that could contribute to overcome the challenges and to help manage the related uncertainties derived from remote work. 

The research was accepted by the scrum master (the fourth author), and the two other authors were invited to lead in the diagnosis and execution of the action research. All team members were warned about the research and they agreed to participate. The action research cycles were taken between March 19Th and May 6Th, 2020.

\subsection{Action planning} \label{sec:actionPlanning}

Many uncertainties came with the pandemic, and the diagnosis phase showed that many decisions needed to be taken to avoid further problems. When Di2win should start remote work? Which tools should the company use in the remote environment? How to effectively coordinate the team to the project goals? How to align customers with the current situation? Many of these questions needed answers quickly, and many decisions were taken fast to define the new way of working.

During the pandemic evaluation, many meetings were conducted to discuss the evolution of  Covid-19 in Recife and all uncertainties regarding the immediate change and the remote work. In these meetings, many insights emerged on how to conduct the remote work and to manage the uncertainties related to it, helping to define the scope of the first research cycle.

The first cycle should promote the operation standardization, and its goal was to organize and establish guidelines and the tools that would be used in the remote environment. %These actions were made to help evaluate the best way of dividing the employees and specify the activities, tools, platforms that should be used and followed by all the employees in the remote work. The main researcher in this study helped on the design and operation of these tasks.

After the results evaluation, the team decided that the second cycle should improve previous results (e.g. guidelines and tools set up) to better coordinate the software development process. The goal was to promote new guidelines, events, and collaboration activities that would improve the remote way of working, and reduce uncertainties related to coordination, communication of the project.
%(e.g. including those related to project distribution, communication, coordination, control, and infrastructure) related to project information and execution process. 

The third and last cycle was conducted aiming on the maintainability of the software development process in the remote environment. The goal was to establish a few more activities to better manage the uncertainties related to code development, project execution and status. 

Each cycle duration details and sprints covered are presented in table \ref{r:tab2}. Sprint 0 has had only four days. It was performed only to assess the scenario of uncertainties and to pre-define how remote work would be conducted. In addition, due to the start of the remote work on March 19th, we decided to end the sprint 0 on March 18th and start the first cycle with the Sprint 1 in the next day.

\begin{table}
\footnotesize
\centering
\caption{Cycles duration and sprints covered by each\protect\footnotemark[1]}
\label{r:tab2}
\begin{tabular}{lll}
\hline
\textbf{Cycle} & \textbf{Date interval}               & Sprints covered                                                 \\ \hline
0              & March 15th to March 18th, 2020 & \begin{tabular}[c]{@{}l@{}}• Sprint 0 \end{tabular} \\ \hline

1              & March 19th to April 3rd, 2020 & \begin{tabular}[c]{@{}l@{}}• Sprint 1\\ • Sprint 2\end{tabular} \\ \hline
2              & April 6th to April 20th, 2020 & \begin{tabular}[c]{@{}l@{}}• Sprint 3\\ • Sprint 2\end{tabular} \\ \hline
3              & April 22nd to May 6th, 2020   & \begin{tabular}[c]{@{}l@{}}• Sprint 5\\ • Sprint 6\end{tabular} \\ \hline
\end{tabular}
\\ \footnotemark[1]\emph{Only workdays were considered. The gaps among cycles are due to weekends and regional holidays}
\end{table}

\subsection{Actions - Cycle  0:} \label{sec:actions0}

This section describes a experiment carried during a few days before the official lockdown started in Brazil: A small sprint to evaluate what could be done with the work activities and to reduce uncertainties related to remote work.

\subsubsection{A1 - Conduct remote experiments with some employees} \label{sec:A1}  The experiments started on March 15th 2020. The experiment consisted in sending a few employees with laptops to work remotely from home, leaving behind at the office the developers, the technical leader, the scrum master and the directors. 

The idea was to reduce the risk of contact among developers, and test team collaboration, communication and productivity. %The team members were divided into two groups physically distant from each other. 
Each day, during these experiments research team evaluated the pandemic situation and the work in progress. On the March 18th, 2020, the pandemic situation on the region became more serious and the directors decided to send the remaining employees who were at the office, the development team, to work from home, including them.

The experiment was helpful to evaluate the remote work feasibility. It was decided to formally conduct the action research aligned with the sprints, here presented as sprints 1 to 6.

\subsection{Actions - Cycle  1:} \label{sec:actions1}

This section describes all actions carried out during the research cycle 1.

\subsubsection{A2 - Establish Di2win collaboration guidelines} \label{sec:A2}

Since the company's staff started to work from home, guidelines were established by the research team, based on the discussion conducted. %These guidelines describes the work rite to be followed by all employees on Microsoft Teams \cite{williams2011scrum+}
% Sugestão de corte:
\begin{comment}
, such as:

\begin{itemize}
    \item Notify that you are starting the project activities in the general channel of the company;
    \item Notify that you are starting the project activities in the specific channel of the project you are working on;
    \item Notify when you need to be away warning absence and its expected duration;
    \item By the time that you enter in the project channel, you should access the team meeting with voice and video when possible, if the meeting has not started yet, you should open it;
    \item You should fulfil the workday regarding lunchtimes and your workload;
    \item Report worked hours at Azure DevOps Boards.
\end{itemize}{}

\end{comment}

These guidelines allow the team to know who is available, to establish communication in the project teams, to avoid any misunderstandings regarding contacting people at an inappropriate time and to reduce uncertainties regarding team members presence. However, these guidelines target only inside communication. Thus, it leaves the customer without visibility and generates misunderstanding regarding team members work hours.

\subsubsection{A3 - Supply the team with the necessary hardware to do the job} \label{sec:A3}
The company provided the team with the necessary equipment to work from home. %, such as desktops computers and headsets, and paid its transportation to employees' homes. 
Hence, all team members could work from home with all necessary infrastructure from the office. This action resulted in a gain in time because all members could work with their machines in their already well-known environment, and the technology uncertainties related to environment configuration and hardware issues were also reduced. %Such action helped the developers since they did not need to configure their development environment in a brand new computer. 

\subsubsection{A4 - Control of the source code} \label{sec:A4}

The research team decided to control the source code in Gitlab \cite{hethey2013gitlab}. Since it has been used for a long time in the teams, this action reduces uncertainties that might arise during an eventual migration of the legacy projects to a new platform. However, any new project that arises while the team is remotely distributed should use the Azure DevOps Repos \cite{rossberg2019overview}, to facilitate the process of continuous integration and deployment. 

\subsubsection{A5 - Provide access to test and production environments} \label{sec:A5}
All servers from Di2win stayed available for the team members to give support, deploy software increments, and monitor the logs. Hence, any team member could access these servers from home. This action revealed the importance of not making the teams dependent on a unique network access point to enter the server. %, making them available for anyone with the credentials.

\subsubsection{A6 - Define an official chat tool for work} \label{sec:A6}
A chat tool was chosen as the official communication channel. Microsoft Teams was chosen the official chat tool.% Before the pandemic, when the job was co-located, the company already had Microsoft Teams as a channel for work, but little attention had been given to it by the team members. 
When the remote work started, the team saw it as an opportunity to be the indispensable ``communication bridge'' for all employees to contact each other. 

% Sugestão de corte:
%This unique channel is not yet a knowledge management tool, but concentrates all exchange messages regarding information about projects, business decisions, work rules, among others. It avoids misunderstanding, information lost and other uncertainties regarding communication and project information and decisions. Besides, the Microsoft Teams centralize all the calls and meetings made by the team members, and with the customers. 

Furthermore, in the past, some project information was treated in WhatsApp\footnote{www.whatsapp.com} by the team members, keeping it difficult to track the information and generating uncertainties about where to find them, but with Microsoft Teams usage all data regarding the projects are available on the team's channels. %However, one of the existing problems is the limitation of using the platform to communicate with some clients: it is only possible through meetings, and it makes some of the stakeholders talk with the team members via WhatsApp.

\subsubsection{A7 - Define an official document storage tool} \label{sec:A7}
The company decided to use One Drive as the official tool to store all the documents regarding the projects, the business strategies, and other stuff. One Drive has been used since the work was co-located, but when it became distributed, the development team was encouraged to use it more, building documents about the current sprint, manuals for environment configuration, and other things. 

As a result, a better knowledge sharing attitude and mindset was established by the team. Besides, the situation when team members are always asking for some kind of information in an informal face-to-face way, was reduced. Nowadays, team members just need to consult the documents on One Drive, reducing uncertainties related to project information. %However, in the first cycle, it was still a work in progress result. Some of the folders and documents from the team were disorganized and the documents had no official owner - generating some rework. To solve this, some members became responsible for specific documents, making the information more consistent.

\subsubsection{A8 - Define a tool for sprint retrospectives} \label{sec:A8}
Many tools and methods to conduct retrospectives were discussed such the examples at \cite{ScrumBook}. The research team was responsible for defining a tool for sprint retrospective. The tool needed to be online and all members should access the board at the same time. For this reason, the tool selected was FunRetro \cite{funretro}, which gives a lot of board options to make a sprint retrospective. In the final of the sprint, the scrum master creates a board in FunRetro \cite{funretro} and shares the link with the developers; all of them write in the board the facts that went well, did not go well and needs to be improved from the past sprint; after this phase, the team members vote for the most important items, they discuss them and generate some actions to improve their environment. 

Finally, all actions are organized by the scrum master and shared in the team's channel on Microsoft Teams. This process allowed actions evaluation for each cycle and helped build new actions to reduce uncertainty for the next cycles. Some of the actions from the Cycle 1's retrospective were useful to cycle 2, and those from cycle 2 were useful for cycle 3.

\subsubsection{A9 - Migrate activities to Azure DevOps} \label{sec:A9}
Before the pandemic, the team used to use Redmine\footnote{https://www.redmine.org/}. %for planning sprints, registering tasks and bugs, reporting working hours, etc. However, the Redmine server was running in a local machine at the office. 
When all team members needed to work remotely, the team decided to migrate all planning activities to Azure DevOps\footnote{https://azure.microsoft.com/en-us/services/devops/}, so the team's sprint backlog, the tasks, and the bugs should be registered there and could be accessed by any team member anywhere. 

This change was quickly accepted by the team, not just because of the planning activities, but also because it reduced the technological uncertainties related to access a management tool, and by the fact that they could enhance their development process with Azure DevOps.

%and by the fact that they were allowed to organize other things in Azure DevOps, such as new repositories and migrate their local artifacts repository to Azure DevOps Artefacts. Even though team members know the capability of Azure DevOps, they had not enough expertise to use all the features available on the platform, because of it some time was spent to learn about its functionalities.

\subsubsection{A10 - Establish a regular delivery pace} \label{sec:A10}
Knowing that the work would be remotely distributed, the scrum master together with the product owner defined the number of days the sprint would have and the regular schedule for releasing new versions of the system. %Since the product backlog was partially built before the pandemic with some stories accepted, the scrum master and the product owner decides that each sprint would take 5 working days.

At the first day of the sprint, the scrum master would be responsible for taking the prioritized and accepted stories from the product backlog, discuss them with the team members and decide the sprint scope. At the end of the sprint, the team would publish the new features in the homologation environment. In parallel, the product owner and the requirements analysts were validating new stories with the customer and refining the backlog with the technical leader

This action was well rated by the client who was satisfied by being able to keep up with the value delivered. This approach also helped increase the development velocity, always focusing on planning, developing, testing and publishing new features in a 5 days interval. 

%In parallel, the product owner and the requirements analysts were validating new stories with the customer and refining the backlog with the technical leader. Helping to reduce uncertainties by keeping the backlog updated, having new features that the developers could work on, and publishing constantly valuable features in the homologation environment.

\subsubsection{A11 - Align expectations with the stakeholder.} \label{sec:A11}

The experience from the experiment, the guidelines, the results, the actions, the environment was shared with a client that had a team working in their premises. This team helped to define the process of working distributed for the client. They managed market uncertainties aligning the client expectations and the possible impact that the transition could cause in the project. This action resulted in some benefits, like keeping constant contact with the client's decisions, influencing their decisions, and making decisions side by side with them.

\subsection{Actions - Cycle 2:} \label{sec:action2}
This section describes all actions carried out during the research cycle 2.

\subsubsection{A12 - Give titles to the meetings in Microsoft Teams} \label{sec:A12}
After defining the collaboration guidelines at cycle 1, and lots of people using Microsoft Teams daily it was quite hard to find important messages sent in the meetings with many other messages thus leading to uncertainties regarding the project information once again.

For the second cycle, a refinement on the collaboration guideline was defined. Titles for each meetings were given to facilitate future searches. This action made the team create meetings with self-explanatory titles. %For instance, when two developers discussed the database model needed for a requirement, the title of the meeting was "Database model for REQ XX", making it easy to find images describing the model that those team members developed and discussed at the meeting chat just by searching for the requirement number. However, the institutionalization was not complete and not all team members applied to this, leading to difficulties on finding project information.

\subsubsection{A13 - Establish a daily meeting in Microsoft Teams at a specific time} \label{sec:A13}

Another action was to carry on daily meetings on the Microsoft Teams, to guarantee the most effective dynamic. Besides, a common and consensus time was established for these meetings. Although this was a common practice for the co-located sprints, it was not until then that the team members come to in a consensus of the online time for the daily meetings. 

%To keep the daily meetings as fast as possible all team members opened the Kanban board on Azure DevOps and followed the usual topics: (i) what they did in the day before, (ii) what they are planning to do, and (iii) if they had any impediments. 

This action brought some good benefits such as avoided former problem of some developers not showing up, unclear environment for the meeting (e.g. Hangout, Skype, Zoom) and untreated impediments related to the development process. It therefore, reduced uncertainties regarding availability of team members to the meetings and clarity on the meeting’s environment, and moment to present and address the impediments from the former day. 

\subsubsection{A14 - Establish a stand-down report} \label{sec:A14}
When co-located, a scrum master can notice an impediment through many clues, such as a look, someone putting their hands on the head, someone skipping lunch, etc. In order to guarantee the productivity, since the work became remote the scrum master missed knowing or noticing problems, issues and impediments. %Moreover, the scrum master would not know how the team members ended their day, and if they had any issues holding them behind. 

A stand-down report, near the end of the day was proposed. All team members should pass their status and share dependencies or issues related to the stories they are working on. Whenever possible, the scrum master asked for the report by videoconference, if not, the team members left a brief message describing it. This action %is not a second daily meeting, it 
is more like a status report in a faster way that helped manage uncertainties related to technology and project tasks. %It resulted in knowing earlier the issues and impediments that team members were facing to solve more effectively those issues. However, sometimes the team members felt they were requested to tell everything they were doing, like if they needed to be accompanied.

\subsubsection{A15 - Conduct training and workshops} \label{sec:A15}
After the first cycle, the scrum master saw the need for training the team in the Azure DevOps aiming to reduce technological uncertainties as some members were showing some difficulty in using it. Hence, the technical leader and the scrum master were responsible for showing how to use Azure DevOps for registering tasks, planning activities, report working hours, and other things. After that, the team members started to use the Azure DevOps regularly, and all their work is now registered on the platform. %Besides that, the scrum master started to do some workshops in the middle of the day about subjects related to our job, like project management. These actions had high acceptance by the team members that felt encouraged to start their workshops and training.

\subsubsection{A16 - Shorten the distance among the team members} \label{sec:A16}
In the first sprints, the scrum master realized that some of the developers were having doubts about the acceptance criteria and the prototype of the stories. To solve these doubts, the developers asked for clarification to the technical leader who was responsible to contact the product owner and the requirements developers asking for an explanation about some of the stories.

This process overwhelmed the technical leader that needed to constantly keep in contact with the product owner. To solve it, the scrum master encouraged the developers themselves to ask the product owner about any issue related to the stories, aiming to reduce the distance between the teams.

%This action made all team members closer, helped in stories specification, and reduced the technical leader overload. However, the team noticed the necessity to improve the stories to reduce doubts.

\subsubsection{A17 - Conduct socialization events} \label{sec:A17}

Encouraged by the scrum master, the developers together with the  technical leader decided to configure a server with an online first-person shooter game. The goal was to reduce socio-human uncertainties and have some social and fun moments in the middle of a pandemic while everybody needs to stay home.

When the company made the server available one team member started to configure the game server. With the server done, %other team members, employees, and the scrum master entered the server and started the game. 
the experience was welcome for everybody that was used to playing games, and made the employees closer and helped the team members strengthen their bonds. However, after few days the team members lost interest on the games. 

\subsubsection{A18 - Establish a code review} \label{sec:A18}

A situation emerged a few hours before the sprint review with the customer. The team members found a lot of small bugs after the deployment in the homologation environment.  This problem demanded the team to develop quick fixes in a small period of time, push them to the repository and published in the server without the proper testing activities. All team members agreed that they should improve code quality in order to avoid this to happen again. 

A code review process was established. %For any feature in which a team member works, another team member needs to make a code review in the developed code to ensure the quality and the patterns used.
More experienced team members were responsible for making the code review of most critical features. This action reduced the uncertainties related to code quality, the quality was improved, fewer bugs were found and no more quick fixes were needed after deploying the system for sprint review. On the other hand, this action overwhelmed the more experienced. %, because they were constantly called to review the features and give feedback about the code the developers made.

\subsection{Actions - Cycle 3:} \label{sec:actions3}
This section describes all actions carried out during the research cycle 3.

\subsubsection{A19 - Build a wiki in Azure DevOps} \label{sec:A19}
When the job was co-located the team members used to ask each other in an informal face to face way about instructions related to the configuration of environments, deployment of the project, installation of modules, for example. This kind of information continues being needed in the remote work, so the first part of it was described in chat messages on Microsoft Teams. Later on, the team decided to create the project Wiki on Azure DevOps, which offers a specific tab for this, making sure that important information regarding instructions about environments, project installation, and deployment were all concentrated in Azure DevOps. This action resulted in an improvement in the knowledge shared and reduced uncertainties regarding knowledge about the project. However, not all information about the project was present in the Wiki, some of it is still being tacit knowledge from the team members.

\subsubsection{A20 - Make rounds of knowledge sharing} \label{sec:A20}
In the last sprint of the cycle 2, the two most experienced developers worked in some challenging and complex stories. However, share knowledge about these challenges were not easy while work from home. %In the co-located environment the team members shared their knowledge through informal face to face communication, but now it was quite hard to do this in the remote work. 
Hence, to align all team members with the knowledge necessary to develop the critical stories, the technical leader suggested that the two developers make a brief explanation about the feature implemented. %, no more than 30 minutes, about the developed feature, not only explaining the feature but also the code developed. 
This action was well-received by the team members that liked the idea and suggested making it more often. Although brief, it helped all team members capable of developing similar critical stories in the next sprints, and also reduce technological uncertainties related to project stories and developers know-how.

\subsubsection{A21 - Define a communication schedule with the stakeholders} \label{sec:A21}
At the end of the cycle 2, a communication schedule with the customer was created. Many meetings were already held, but the communication schedule helped the client's project manager to visualize who participates in each meeting, the frequency of the meeting and in which channel the communication takes place. This action resulted in reducing uncertainties regarding project progress, and helped the team members and the client to understand each communication event that happens in the project development.

\subsubsection{A22 - Regular feedbacks for team working from home} \label{sec:A22}

%The team members were forced by surprise to switch to a remote way of working, while many of them had less than 2 years of experience and had never experienced a remote distributed work environment. Many uncertainties were regarding this situation, and as seen in the study of Paul Ralph \textit{et al.,} \cite{Ralph2020Pandemic} the home office ergonomics, the distractions caused by people with whom the team members live, the absence of fitness facilities and the fear of pandemic can truly affect well being and productivity of the team members. 
Although trying to reduce the effects that those challenges could bring, the scrum master decided to schedule regular feedback from the team members about their work from home, aiming to detect signs of possible issues that could arise. During the feedbacks, the team members argued that they were missing office ergonomics and comfort, especially the chair on which they used to sit. To overcome this situation and to reduce the uncertainties related to it the scrum master and team are studying the possibility of taking the office furniture of each team member to their homes.

\subsubsection{A23 - Game rounds in work hour} \label{sec:A23}
As we saw in A17 \ref{sec:A17}, the game server configured by the team was forgotten a few days after due to a loss of interest by the team members. Despite most of this loss was because the team members were busy with personal matters outside the work hours, they continued to have the intent to play together.

To change the approach taken before, the scrum master suggested the team members play in the middle of the day, after lunch for just 30 minutes. At the end of sprint 6, the team tried it and the experience was well received, almost all team members engaged in the game. %This new approach for gaming during the work hours helped the team to de-stress after a hard sprint.

\subsection{Evaluation and Analysis} \label{evaluation}
The section discusses the study results and the impact that the actions made in the project development process, the relationship of team members, the quality of the code, and the use of the new platforms defined. The graphics ploted aims to present some of the benefits of the actions, and it were built using the open source platform metabase \footnote{www.metabase.com} and Microsoft Excel.

The project developed during the study started on December 2019. Although some critical business requirements changed in February 2020 and a new phase needed to be started. This phase started a few weeks before the spread of Covid-19 intensification in Recife. Therefore, the pandemic practically forced the development of this new phase of the project to be full remotely.

During the action research cycles, almost all critical requirements were implemented, only a small group of requirements still remained for other sprints, to be refined and evaluated by the customer. Although it is still to be finished, the customer and some stakeholders gave good feedback about the progress of the project during the quarantine and felt satisfied with the results.

\begin{figure}
    \centering
    \subfigure[]{\includegraphics[width=0.49\textwidth]{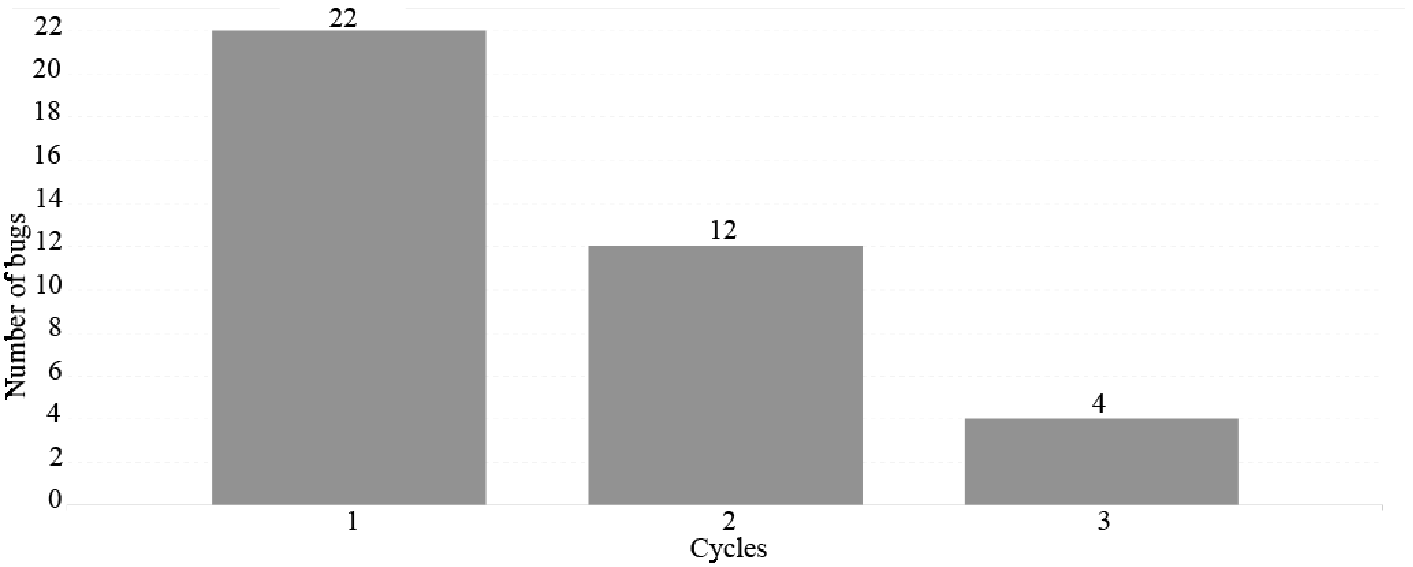}} 
    \subfigure[]{\includegraphics[width=0.49\textwidth]{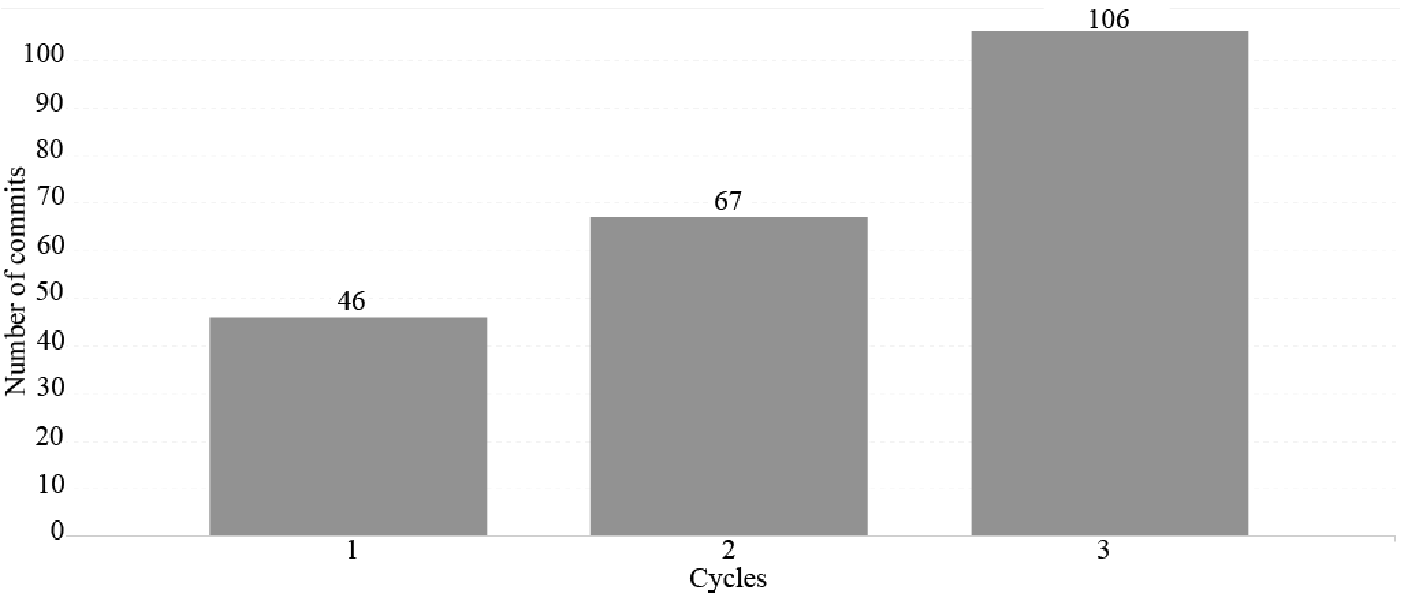}} 
    \caption{(a) Number of bugs per cycle (b) Number of commits per cycle in the most critical module of the project}
    \label{figure:fig1}
\end{figure}

Many benefits were taken from the actions presented, and it reflected directly in project quality and coordination, and in a better collaborative environment for all team members. Among them, the quality of the features developed \cite{oliveiracode} is presented in Figure \ref{figure:fig1}(a), as we can see the number of bugs found in the project was reduced over the cycles, consequently reducing the uncertainties around technology and quality.
\begin{comment}

\begin{figure}
    \centering
    \includegraphics[scale=0.5]{Fig1.eps}
    \caption{\label{figure:fig1}  Number of bugs per cycle}
\end{figure}
\end{comment}

In cycle one, the customer evaluation found 22 bugs in the first sprint, then this number reduced to 12 in cycle 2, and finally, in the last cycle, only 4 bugs were found by the team members and/or the customer. It is worth mentioning that the process to evaluate and approve the features developed continued to be done by the customer during each sprint of each cycle in the sprint review. Moreover, the developers continued their activities of cross-testing, when a developer is responsible for testing the feature that another partner has developed, intending to find bugs and improve the quality of the code. 

Another point that corroborates the reduction of the bugs is the fact that the number of commits in one of the most critical modules of the project grew while the number of bugs diminished, as shown in Figure \ref{figure:fig1}(b). This improvement in the code quality can be related to some actions like A18, that established code review process. During these code reviews the technical leader and the P4  developer tested and gave constant feedback, helping the developers fix any problem quickly, improving their code, and reducing uncertainties related to technology.

\begin{comment}
\begin{figure}
    \centering
    \includegraphics[scale=0.5]{Fig2.eps}
    \caption{\label{figure:fig2}  Number of commits per cycle in the most critical module of the project}
\end{figure} 
\end{comment}

Another action related to the reduction of bugs is the A16, which shortened the distance among all the team members, making their work closer, better understanding each other points, and managing uncertainties regarding requirements specification. 

Moreover, an action that may be corroborated to bug reduction is the A20, that improved knowledge sharing between developers and helped them all to achieve skills for developing complex features.

\begin{figure}
    \centering
    \subfigure[]{\includegraphics[width=0.49\textwidth]{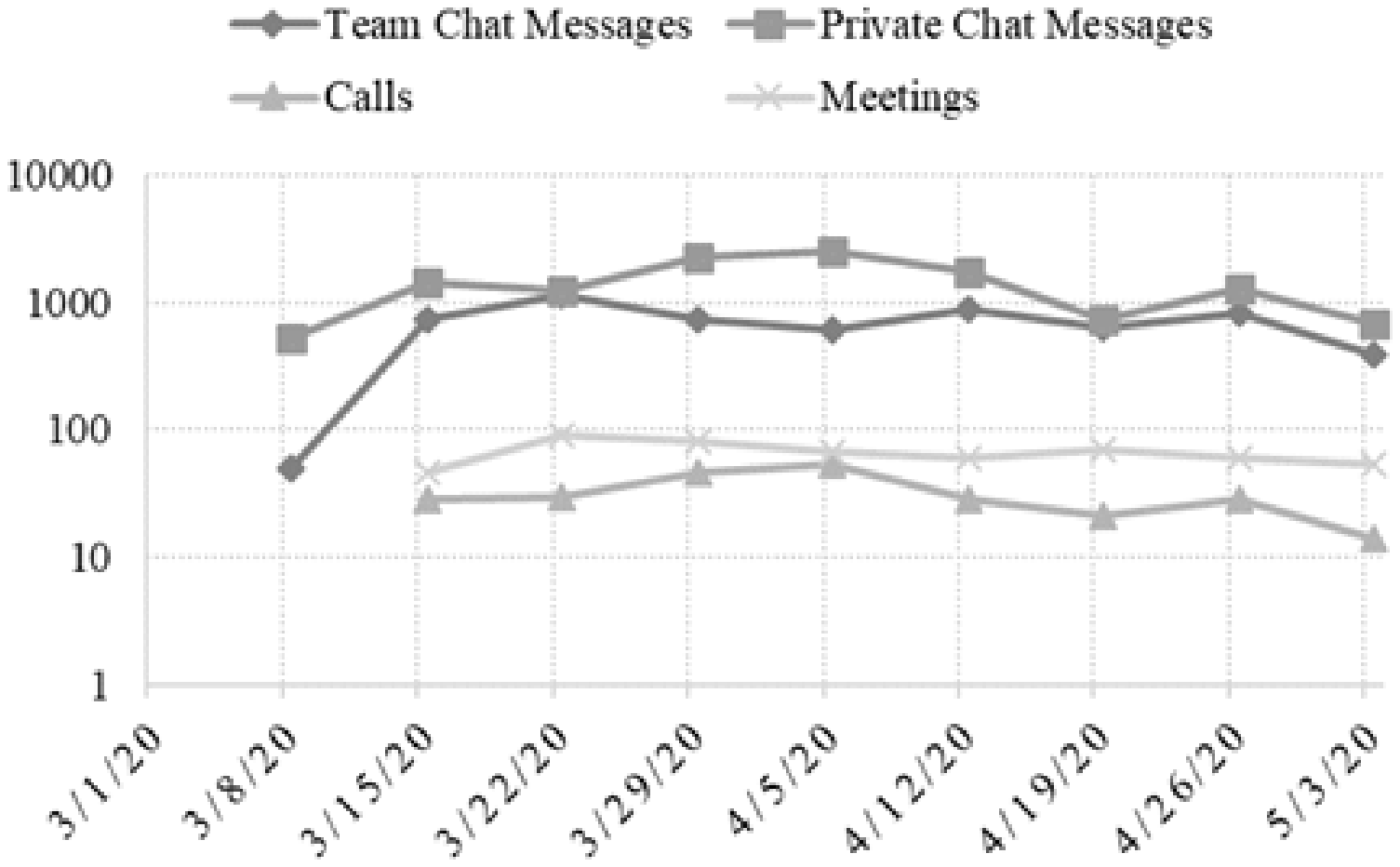}} 
    \subfigure[]{\includegraphics[width=0.49\textwidth]{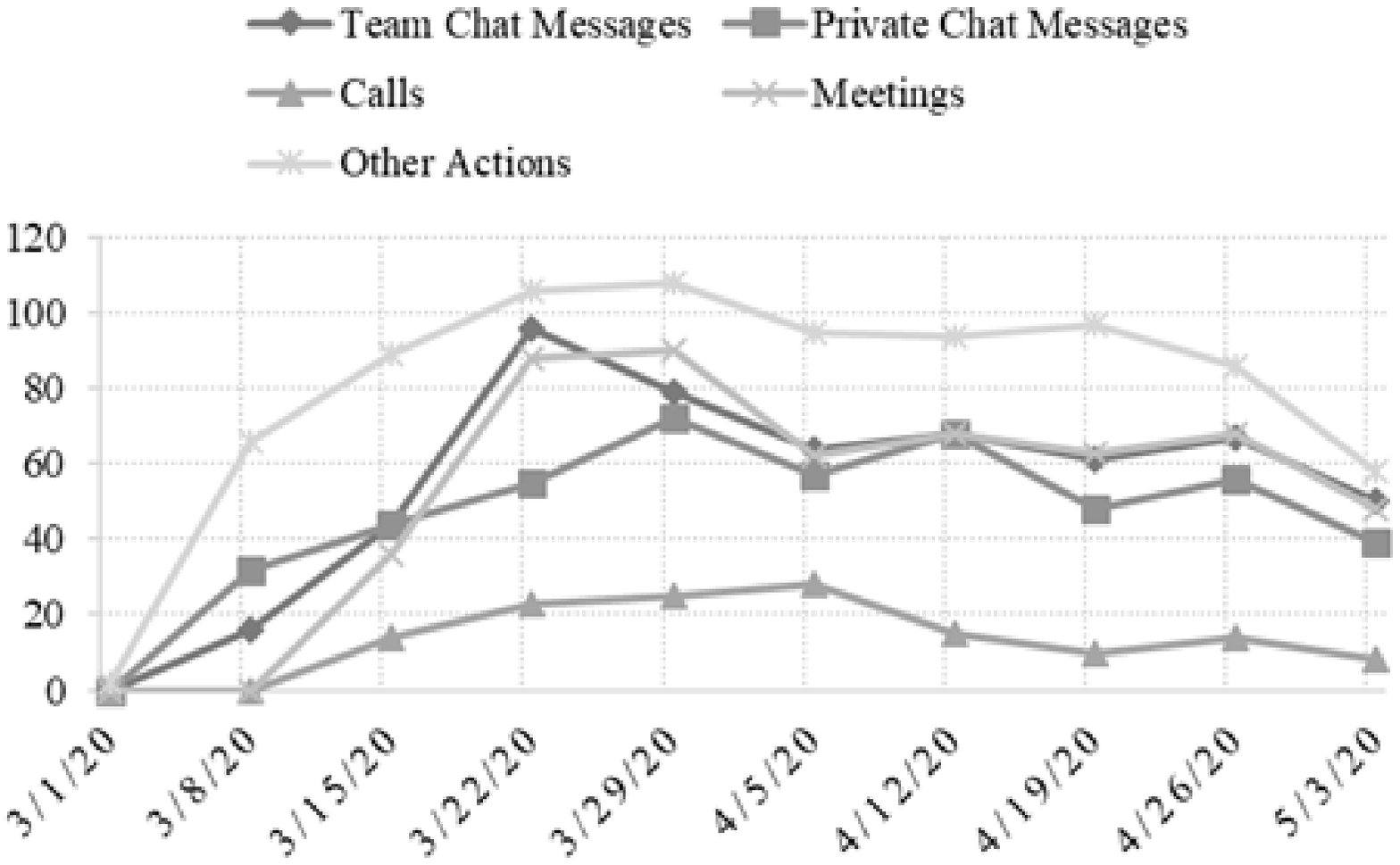}} 
    \caption{(a) Number of activities by type and per week (b) Number of users interactions by activity type and per week}
    \label{figure:fig2}
\end{figure}

The actions related to defining tools and guidelines for the remote work, such as A7, A8, and especially A2 and A6, that defines the communication tool for the remote work and collaboration guidelines show its importance. Therefore, it establish a central and unique chat tool for all the communication, aiming to concentrate all the information regarding the company operation and to help knowing if a collaborator is available online or not. Before the pandemic, little attention was given to Microsoft Teams, most of the project information was exchanged through informal face-to-face communication or through other channels, and only a few members used the tool to contact employees outside the office. During the pandemic, the use of Microsoft teams expanded, as we can see in Figure \ref{figure:fig2}(a), the number of activities was so low at the beginning of March, 2020 with only a few team and private chat messages exchanged and no meeting and calls made. However, in the week of March 15th, 2020, when work started to be remote, the number of team and private chat messages, calls and meetings grew considerably, registering more than 85 meetings, 1500 private chat messages, 50 calls, and 600 team chat messages during some weeks. Since values from these measurements have distinct order of magnitudes, a logarithmic scale was used to visualize all data in one graph.

\begin{comment}
\begin{figure}
    \centering
    \includegraphics[scale=0.5]{Fig3.eps}
    \caption{\label{figure:fig3} Number of activities by type and per week}
\end{figure}
\end{comment}

The action of defining a communication tool showed the importance to concentrate all the information exchanged about the projects in one environment, aiming to replace the old exchanges of information in the co-located environment through informal face-to-face communication and to reduce uncertainties related to project information, presence of employees and project meetings. Another data that corroborates the fact that Microsoft Teams was adopted by the team members are shown in Figure \ref{figure:fig2}(b), as we can see the sum of interactions among users by activity type and per week has grown together with the number of activities. It shows that, at the beginning of March, no meetings or calls were made by the team members. However, during the pandemic, we had more than 80 meetings and team chat messages interactions per week in the platform, and more than 50 and 20 interactions among users per week using respectively private chat messages and calls. The calculation used in this graph is based on interactions among users, e.g whether the scrum master makes a meeting with a developer and then makes a meeting with the product owner in the same week, it will count two interactions. The reduction of activities in the week of April 12Th, 2020, is due to some regional and national holidays, and also because of one team member who went on vacation.

\begin{comment}
\begin{figure}
    \centering
    \includegraphics[scale=0.5]{Fig4.eps}
    \caption{\label{figure:fig4} Number of users interactions by activity type and per week}
\end{figure} 
\end{comment}

\subsection{Reflections and Learning} \label{sec:reflection}

The diagnosis phase showed how challenging the pandemic would be to Di2win, with many uncertainties regarding the future of the startup, its customers and its software development process. During the action research, the actions provided good results, insights, and lessons learned about how to manage the uncertainties related to the remote work during the spread of Covid-19. The actions assisted the team to improve the quality of the code, to understand project requirements, events schedule, in the relationship among the members, the knowledge sharing, and consequently the feedback from the customer about the work done. 

In the final cycle, the technical leader was responsible to give feedback to the developers about their behavior and performance during the remote work. However, not only the evaluation of each developer was considered, but the technical leader also asked all developers about the study and its actions. The developers agreed that the actions brought many benefits to the project and contributed to: help the team achieve technical skills, learn more about the project business rules, enhance their development process, reduce the technology uncertainties, closing the distance between the team members and improve the quality of the features developed.

During the feedback sessions with the developers, the P5 said: ``the knowledge sharing round was quite helpful, we should do this frequently, it clarified the uncertainties that I had about the implementation of the complex requirements in the backend''. Then, the P4 stated ``It is interesting that we were able to establish the code review policy during remote work, for me, it was one of the most important things. It put the team on a new level and made me feel that we produced code with more quality and fewer bugs''. P2 also said that ``I think we gave a big step when we started to make more contact with all the team members, many doubts about the business domain were clarified and misunderstandings were not more frequent'', he also said that ``The game rounds at the end of the sprint 6 were very helpful to de-stress me, I felt better to work.''. Furthermore, P3 told us ``I think we should keep doing our job in this way, with all these actions, it became really clear how our process work and what we should do''. He also said that ``The virtual meetings with the entire development team makes me feel like I am in the office, talking with my work colleagues about the project and sometimes just chatting about random things.''

The developers' statements highlight some of the lessons learned from the action research. Furthermore, the actions taken answer the research question by showing that to manage and reduce uncertainties during the pandemic, it was necessary to promote socialization events, establish socialization guidelines, knowledge sharing rounds, establish code development standards, shorten the distance among team members, provide access to production and test servers, define tools for different purposes, set a schedule of project events and meetings, conduct training and workshops and maintain constant contact with the customers'.

\subsection{Threats to validity} \label{validity}
Besides the fact that we followed the steps of the action research described in the method section, some threats to validity should be considered about the study.

\subsubsection{Construct validity} \label{construct} It defines in what degree the used measures that are studied, really represent what the researchers intended to look for and what is investigated according to the research questions \cite{Runeson2008GuidelinesFC}. The most important threat related to this validity is the fact that the first and fourth authors work in the project as the technical leader and scrum master respectively, which can make them tend to observe the results of the actions and interpret them based on their values and expectations. However, to mitigate this threat, the external researchers were responsible for constantly evaluate the results of the actions and verify them based on the data available. Whether an issue was perceived by the external researchers about any consideration from the internal researcher, they discussed the issue together to resolve it with further analysis. 

\subsubsection{Internal validity} \label{internal} It aims to ensure that the results are derived and based on the data, and it also focus on the study design. In our action research, all information from all cycles was captured through the team observation, the feedback sessions, and the platforms that the team is using, such as Azure DevOps, Microsoft Teams, and source code repositories. To ensure the internal validity all the data were reviewed by the authors and the team members were consulted.

\subsubsection{External validity} \label{external} This threat is related to the capacity to generalize the results of the study. An external threat could be related to the fact that only a single project has been studied with a focus on a single team from the startup. However, we believe that the approach used was systematic, and any startup or company in a similar situation can use the proposed actions.

% Sugestão de corte:
%\subsubsection{Conclusion validity} \label{conclusion} It intends to guarantee that the results are consistent and can be replicated following the same steps. To ensure this validity, all the quantitative data from the study was stored in a database, organized in some tables and the results were published in a private cloud platform. Furthermore, all qualitative data analysis is stored in cloud platforms for any need for future access. Finally, the data analysis process was done ensuring that the data would only be normalized for research, and to guarantee it, all the researchers and the scrum master reviewed the data extracted from the platforms.

\section{Conclusion}  \label{sec:conclusions}
The Covid-19 pandemic surprised the world, forcing companies to quickly change their ways of working. Companies that used to have their activities in a co-located way faced many uncertainties migrating their operation to the remote way of working, uncertainties related to technology, market, socioeconomic challenges, infrastructure, and many others. Managing uncertainties together with the adaptation of some agile practices can be a determining factor to maintain productivity during the remote work in the pandemic.

This study presents the results of an action research between the period of March 15th to May 6th, 2020, conducted in a medium size software Startup during the quarantine in Recife due to the Covid-19 pandemic.

The results from the actions taken during the study can be helpful for other startups and companies that are facing the consequences of the pandemic and the uncertainties related to it. However, the study also provides the academic community with manners for dealing with uncertainties in remote teams that are not used to work remotely. Besides it, the main contribution of this study is the lessons learned presented, as well as an action-research' steps. Consequently, aligning the theory with the practice. 

In this paper, we focused on software startup. However, these findings can be applied in a different company utilizing agile methods to improve remote works. %Furthermore, the action research put forward lessons learned that can help agile software development teams improve their way of work in a remote environment.
Furthemore, Global software development (GSD) usually suffers from these uncertainties \cite{Uncertainty&GTM2018}. However, our study aims to take action to manage uncertainties in the context of startups that suddenly started to work from home.

As future work, it is expected to execute the actions taken with other teams and projects at the company, and hopefully in other startup environments. Furthermore, we will develop a model based on the action results that guides companies through the needed changes for migration to remote work.

\bibliographystyle{IEEEtran}
\bibliography{springerart}

\vspace{2cm}

\end{document}